\documentclass[aps,prd,twocolumn,showpacs,preprintnumbers,superscriptaddress,amsmath,amssymb]{revtex4}
\usepackage{graphicx}
\usepackage{dcolumn}
\usepackage{color}
\usepackage{epsfig} 
\graphicspath{{ps}}

\newcommand{\gev} {\ensuremath{\, {\mathrm{GeV}}     }}
\newcommand{\mev} {\ensuremath{\, {\mathrm{MeV}}     }}
\newcommand{\gevc}{\ensuremath{\, {\mathrm{GeV}/c^2} }}
\newcommand{\mevc}{\ensuremath{\, {\mathrm{MeV}/c^2} }}

\newcommand{\ecm} {\ensuremath{ E_{\mathrm{c.m.}} }}
\newcommand{\gisr}{\ensuremath{ \gamma_{ISR} }}
\newcommand{\sqs} {\ensuremath{ \sqrt{s} }}

\newcommand{\RMS} {\ensuremath{ M^2_{\mathrm{rec}} }}

\newcommand{\dd}  {\ensuremath{ D \overline D }}
\newcommand{\dpdm}{\ensuremath{ D^+ D^- }}
\newcommand{\ddb} {\ensuremath{ D^0 \overline D{}^0    }}
\newcommand{\ddst}{\ensuremath{ D^0 \overline D{}^{*0} }}

\newcommand{\ps}  {\ensuremath{ \psi(3770) }}
\newcommand{\psiddt} {\ensuremath{\psi(4415)\to D \overline D{}^{*}_2(2460) }}

\newcommand{\md}   {\ensuremath{ M_{D} }}
\newcommand{\mdb}  {\ensuremath{ M_{\overline D} }}
\newcommand{\mdd}  {\ensuremath{ M_{D \overline D} }}
\newcommand{\mdpdm}{\ensuremath{ M_{D^+ D^-} }}
\newcommand{\mddb} {\ensuremath{ M_{D^0 \overline D{}^0} }}

\newcommand{\ee}    {\ensuremath{ e^+e^- }}
\newcommand{\eedd}  {\ensuremath{ e^+e^- \to D \overline D }}
\newcommand{\eeddg} {\ensuremath{ e^+e^- \to D \overline D \gamma_{ISR}}}
\newcommand{\eedpdm}{\ensuremath{ e^+e^- \to D^+ D^- }}
\newcommand{\eeddb} {\ensuremath{ e^+e^- \to D^0 \overline D{}^0 }}

\newcommand{\eeddp}     {\ensuremath{ e^+e^- \to D^0 D^- \pi^+ }}

\newcommand{\eeddnpig}{\ensuremath{ e^+e^- \to D \overline D (n)\pi \gisr  }}
\newcommand{\eeddpi}  {\ensuremath{ e^+e^- \to D \overline D \pi^0_{\mathrm{miss}} \gamma_{ISR}}}
\newcommand{\eeddstg} {\ensuremath{ e^+e^- \to D \overline D{}^{*} \gisr }}
\newcommand{\eeddstng} {\ensuremath{ e^+e^- \to D^0 \overline D{}^{*0} \gisr }}
\newcommand{\misid}   {\ensuremath{ e^+e^- \to D \overline D \pi^0 }}
\newcommand{\eeddmpi} {\ensuremath{ e^+e^- \to D^0 D^-\pi^+_{\mathrm{miss}} \gamma_{ISR}}}

\newcommand{\dpi} {\ensuremath{ \overline D{}^{*}  \to \overline D\pi^0_{\mathrm{miss}} }}

\newcommand{\dg}  {\ensuremath{ \overline D{}^{*0} \to \overline D{}^{0}\gamma }}

\newcommand{\seedd}  {\ensuremath{\sigma( e^+e^- \to D \overline D ) }}
\newcommand{\seedpdm}{\ensuremath{\sigma( e^+e^- \to D^+ D^- )}}
\newcommand{\seeddb} {\ensuremath{\sigma( e^+e^- \to D^0 \overline D{}^0 ) }}

\newcommand{\seedstdst}{\ensuremath{\sigma( e^+e^- \to D^{*+} D^{*-} )}}
\newcommand{\seeddst}  {\ensuremath{\sigma( e^+e^- \to D^{\pm} D^{*\mp} )}}

\begin{document}

\title{\quad\\[0.5cm] Measurement of the near-threshold $e^+e^- \to D \overline D $ cross section using initial-state radiation }.

\affiliation{Budker Institute of Nuclear Physics, Novosibirsk}
\affiliation{Chiba University, Chiba}
\affiliation{University of Cincinnati, Cincinnati, Ohio 45221}
\affiliation{Justus-Liebig-Universit\"at Gie\ss{}en, Gie\ss{}en}
\affiliation{The Graduate University for Advanced Studies, Hayama}
\affiliation{Gyeongsang National University, Chinju}
\affiliation{Hanyang University, Seoul}
\affiliation{University of Hawaii, Honolulu, Hawaii 96822}
\affiliation{High Energy Accelerator Research Organization (KEK), Tsukuba}
\affiliation{University of Illinois at Urbana-Champaign, Urbana, Illinois 61801}
\affiliation{Institute of High Energy Physics, Chinese Academy of Sciences, Beijing}
\affiliation{Institute of High Energy Physics, Vienna}
\affiliation{Institute of High Energy Physics, Protvino}
\affiliation{Institute for Theoretical and Experimental Physics, Moscow}
\affiliation{J. Stefan Institute, Ljubljana}
\affiliation{Kanagawa University, Yokohama}
\affiliation{Kyungpook National University, Taegu}
\affiliation{\'Ecole Polytechnique F\'ed\'erale de Lausanne (EPFL), Lausanne}
\affiliation{University of Ljubljana, Ljubljana}
\affiliation{University of Maribor, Maribor}
\affiliation{University of Melbourne, School of Physics, Victoria 3010}
\affiliation{Nagoya University, Nagoya}
\affiliation{Nara Women's University, Nara}
\affiliation{National Central University, Chung-li}
\affiliation{National United University, Miao Li}
\affiliation{Department of Physics, National Taiwan University, Taipei}
\affiliation{H. Niewodniczanski Institute of Nuclear Physics, Krakow}
\affiliation{Nippon Dental University, Niigata}
\affiliation{Niigata University, Niigata}
\affiliation{University of Nova Gorica, Nova Gorica}
\affiliation{Osaka City University, Osaka}
\affiliation{Osaka University, Osaka}
\affiliation{Panjab University, Chandigarh}
\affiliation{RIKEN BNL Research Center, Upton, New York 11973}
\affiliation{Saga University, Saga}
\affiliation{University of Science and Technology of China, Hefei}
\affiliation{Seoul National University, Seoul}
\affiliation{Sungkyunkwan University, Suwon}
\affiliation{University of Sydney, Sydney, New South Wales}
\affiliation{Toho University, Funabashi}
\affiliation{Tohoku Gakuin University, Tagajo}
\affiliation{Department of Physics, University of Tokyo, Tokyo}
\affiliation{Tokyo Institute of Technology, Tokyo}
\affiliation{Tokyo Metropolitan University, Tokyo}
\affiliation{Tokyo University of Agriculture and Technology, Tokyo}
\affiliation{Virginia Polytechnic Institute and State University, Blacksburg, Virginia 24061}
\affiliation{Yonsei University, Seoul}
 \author{G.~Pakhlova}\affiliation{Institute for Theoretical and Experimental Physics, Moscow} 
  \author{I.~Adachi}\affiliation{High Energy Accelerator Research Organization (KEK), Tsukuba} 
  \author{H.~Aihara}\affiliation{Department of Physics, University of Tokyo, Tokyo} 
  \author{K.~Arinstein}\affiliation{Budker Institute of Nuclear Physics, Novosibirsk} 
  \author{V.~Aulchenko}\affiliation{Budker Institute of Nuclear Physics, Novosibirsk} 
  \author{T.~Aushev}\affiliation{\'Ecole Polytechnique F\'ed\'erale de Lausanne (EPFL), Lausanne}\affiliation{Institute for Theoretical and Experimental Physics, Moscow} 
  \author{A.~M.~Bakich}\affiliation{University of Sydney, Sydney, New South Wales} 
  \author{V.~Balagura}\affiliation{Institute for Theoretical and Experimental Physics, Moscow} 
  \author{A.~Bay}\affiliation{\'Ecole Polytechnique F\'ed\'erale de Lausanne (EPFL), Lausanne} 
  \author{I.~Bedny}\affiliation{Budker Institute of Nuclear Physics, Novosibirsk} 
% ...
\collaboration{The Belle Collaboration}

\begin{abstract}
We report measurements of the exclusive cross section for $e^+e^- \to
D \overline D $, where $D=D^0$ or $D^+$, in the center-of-mass energy
range from the $D \overline D $ threshold to 5 \gev\ with
initial-state radiation. The analysis is based on a data sample
collected with the Belle detector with an integrated luminosity of
$673$ $\mathrm{fb}^{-1}$.
\end{abstract}


\maketitle
\tighten
{\renewcommand{\thefootnote}{\fnsymbol{footnote}}}
\setcounter{footnote}{0}

\noindent

The total cross section for hadron production in \ee\ annihilation in
the \sqs\ region above the open-charm threshold was measured precisely
by the Crystal Ball~\cite{cb:cs} and BES~\cite{bes:cs}
collaborations. However, the parameters of the $J^{PC}=1^{--}$
charmonium states obtained from fits to the inclusive cross
section~\cite{seth,bes:fit} are poorly understood
theoretically~\cite{barnes}. Since interference between different
resonant structures depends upon the specific final states, studies of
{\it exclusive} cross sections for charmed meson pairs in this energy
range are needed to clarify the situation.  Recently, CLEO-c performed
a scan over the energy range from 3.970 to 4.260\gev\ and measured
exclusive cross sections for $D \overline D$, $D \overline {D}{}^*$
and $ D^* \overline {D}{}^*$ final states at twelve points with high
accuracy~\cite{cleo:cs}. Belle has used a partial reconstruction
technique to perform first measurements of the exclusive cross
sections \seeddst\ and \seedstdst\ for \sqs\ near the $D^{+}D^{*-}$
and $D^{*+}D^{*-}$ thresholds with initial-state radiation
(ISR)~\cite{belle:dst}. Recently Belle~\cite{belle:psi} has reported a
measurement of the exclusive cross section for the process \eeddp\ and
the first observation of \psiddt\ decay.

In this paper we report measurements of the exclusive cross sections
for the processes \eedpdm\ and \eeddb\ using ISR that are a
continuation of our studies of the near-threshold exclusive open charm
production. Recently several new charmonium-like states were observed
in this mass range ($Y(4260)$~\cite{babar:4260,belle:4260}, $Y(4360)$,
$Y(4660)$~\cite{belle:4360}, $X(4160)$~\cite{belle:4160}) decaying to
either open- or closed-charm final states. Our study provides further
information on the dynamics of charm quarks at these center of mass
energies.  The data sample corresponds to an integrated luminosity of
$673\,\mathrm{fb}^{-1}$ collected with the Belle detector~\cite{det}
at the $\Upsilon(4S)$ resonance and nearby continuum at the KEKB
asymmetric-energy \ee\ collider~\cite{kekb}.

We select \eeddg\ signal events by reconstructing both the $D$ and
$\overline D$ mesons, where $\dd=\ddb$ or \dpdm. In general, the
\gisr\ is not required to be detected; its presence in the event is
inferred from a peak at zero in the spectrum of the recoil mass
against the \dd\ system. The square of the recoil mass is defined as:
\begin{eqnarray}
\RMS(\dd)=(\ecm - E_{\dd})^2 - p^2_{\dd} ,
\end{eqnarray}
where \ecm\ is the initial \ee\ center-of-mass ($\mathrm{c.m.}$)
energy, $E_{\dd}$ and $p_{\dd}$ are the $\mathrm{c.m.}$ energy and
momentum of the \dd\ combination, respectively.  To suppress
backgrounds we consider two cases: (1) the \gisr\ is out of detector
acceptance in which case the polar angle for the \dd\ combination in
the $\mathrm{c.m.}$ frame is required to be
$|\mathrm{cos}(\theta_{\dd})|>0.9$; (2) the fast \gisr\ is within the
detector acceptance ($|\mathrm{cos}(\theta_{\dd})|<0.9$), in this case
the \gisr\ is required to be detected and the mass of the $\dd\gisr$
combination is required to be greater than $\ecm -0.58\gevc$.  To
suppress background from \eeddnpig\ processes we exclude events that
contain additional charged tracks that are not used in the $D$ or
$\overline D$ reconstruction.

We ensure that all charged tracks originate from the interaction point
(IP) with the requirements $dr<2 \, {\mathrm{cm}}$ and
$|dz|<4\,{\mathrm{cm}}$, where $dr$ and $|dz|$ are the impact
parameters perpendicular to and along the beam direction with respect
to the IP. Charged kaons are required to have a ratio of particle
identification likelihoods, $\mathcal{P}_K = \mathcal{L}_K /
(\mathcal{L}_K + \mathcal{L}_\pi)$~\cite{nim}, larger than 0.6. No
identification requirements are applied for pion candidates. $K^0_S$
candidates are reconstructed from $\pi^+ \pi^-$ pairs with an
invariant mass within $10\mevc$ of the nominal $K^0_S$ mass. The
distance between the two pion tracks at the $K^0_S$ vertex must be
less than $1\,\mathrm{cm}$, the transverse flight distance from the
interaction point is required to be greater than $0.1\,\mathrm{cm}$,
and the angle between the $K^0_S$ momentum direction and the flight
direction in the $x-y$ plane should be smaller than
$0.1\,\mathrm{rad}$. The pion pair candidates are refitted to the
$K^0_S$ mass. Photons are reconstructed in the electromagnetic
calorimeter as showers with energies greater than $50 \mev$ that are
not associated with charged tracks.  Pairs of photons are combined to
form $\pi^0$ candidates. If the mass of a $\gamma \gamma$ pair lies
within $15\mevc$ of the nominal $\pi^0$ mass, the pair is fit with a
$\pi^0$ mass constraint and considered as a $\pi^0$ candidate.  $D^0$
candidates~\cite{foot} are reconstructed using five decay modes: $K^-
\pi^+$, $K^- K^+$, $K^- \pi^- \pi^+ \pi^+$, $K^0_S \pi^+\pi^-$ and
$K^- \pi^+ \pi^0$.  $D^+$ candidates are reconstructed using the decay
modes $K^0_S\pi^+$ and $K^- \pi^+ \pi^+$.  A $\pm 15\mevc$ mass window
is used for all modes except for $K^- \pi^- \pi^+ \pi^+$, where a $\pm
10\mevc$ requirement is applied ($\sim 2.5\,\sigma$ in each case).  To
improve the momentum resolution of $D$ meson candidates, final tracks
are fitted to a common vertex and a $D^0$ or $D^+$ mass constraint is
applied. The $D$ candidates from a sideband region are refitted to the
mass, corresponding to the center of the sideband region with the same
width as the signal window.

\begin{figure}[htb]
\begin{tabular}{cc}
\hspace*{-0.025\textwidth}
\includegraphics[width=0.50\textwidth]{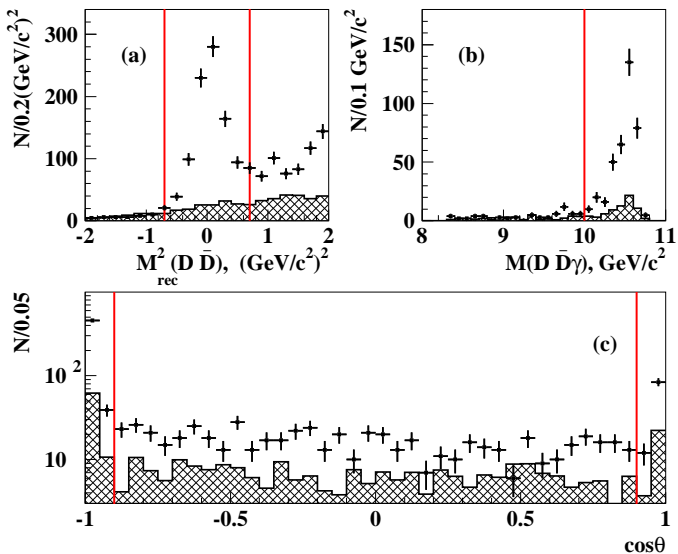}
\end{tabular}
\caption{a) The observed distributions of (a) $\RMS(\dd)$; (b)
  $M(\dd\gisr)$ and (c)\dd\ polar angles. Histograms show the
  normalized contributions from \md\ and \mdb\ sidebands. The selected
  signal windows are illustrated by vertical lines.}
\label{fig1}
\end{figure}

\begin{figure}[htb]
\begin{tabular}{cc}
\hspace*{-0.025\textwidth}
\includegraphics[width=0.5\textwidth]{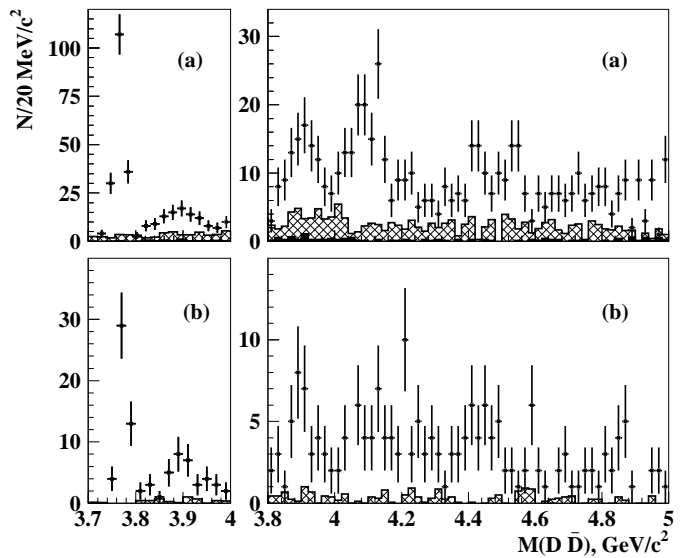}
\end{tabular}
\caption{The mass spectra of \dd\ combinations (points with error
  bars): (a) \ddb; (b) \dpdm.  The total contribution from
  combinatorial background (1--2) is shown as the hatched histogram,
  the contribution from background (4) is shown as the (barely
  visible) solid histogram.}
\label{fig2}
\end{figure}

The distribution of $\RMS(\dd)$ after all the requirements is shown in
Fig.~\ref{fig1}\,(a). A clear peak corresponding to the
\eeddg\ process is evident around zero.  The shoulder at positive
values is due to $e^+ e^- \to D^{(*)} \overline D{}^{(*)} (n)\pi^0
\gisr$ events.  To suppress the tail of such events we define a signal
region by the requirement: $|\RMS(\dd)|< 0.7(\gevc)^2$.  The polar
angle distribution for \dd\ after the requirements on $\RMS(\dd)$ is
shown in Fig.~\ref{fig1}\,(c). The sharp peaking at $\pm 1$ is
characteristic of ISR production and agrees with the Monte Carlo (MC)
simulation. The mass distribution of the $\dd\gisr$ combinations where
the \gisr\ is detected is shown in Fig.~\ref{fig1}\,(b) after all the
selection requirements. It peaks at the \ecm.  The asymmetric shape of
the $M_{\dd\gisr}$ distribution is due to higher-order ISR processes.
The \mddb\ and \mdpdm\ spectra obtained after all the requirements are
shown in Fig.~\ref{fig2}.

The following sources of background are considered:
\begin{itemize}
\item[(1)] combinatorial $D$($\overline D$) mesons combined with a
  real $\overline D$($D$) coming from the signal or other processes;
\item[(2)] both $D$ and $\overline D$ are combinatorial;
\item[(3)] reflection from the processes \eeddpi\ and \eeddstg,
  followed by \dpi, with an extra $\pi^0_{\mathrm{miss}}$ in the final
  state;
\item[(4)] reflection from the process \eeddstg, followed by \dg, with
  an extra soft $\gamma$ in the final state;
\item[(5)] a contribution from \misid\ when an energetic $\pi^0$ is
  misidentified as a single \gisr.
\end{itemize}
The contribution from background (1) is extracted using \md\ and
\mdb\ sidebands that are four times as large as the signal region.
These sidebands are shifted by $30\mevc$ ($20\mevc$ for the $D^0\to
K^-\pi^-\pi^+\pi^+$ mode) from the signal region to avoid signal
over-subtraction. Background (2) is present in both the \md\ and
\mdb\ sidebands and is, thus, subtracted twice.  To account for this
over-subtraction we use a 2-dimensional sideband region, where events
are selected from both the \md\ and the \mdb\ sidebands.  The total
contribution of the combinatorial backgrounds (1--2) is shown in
Fig.~\ref{fig2} as a hatched histogram.  Backgrounds (3--4) are
suppressed by the tight requirement on $\RMS(\dd)$.  The remaining
contribution from background (3) is estimated directly from the data
by applying a similar full reconstruction method to the
isospin-conjugate process \eeddmpi. Here the requirement on absence of
additional charged tracks in the event is relaxed. Since there is a
charge imbalance in the $D^0 D^-$ final state, only events with an
extra missing $\pi^+_{\mathrm{miss}}$ can contribute to the
$\RMS(\dd)$ signal window. To extract the level of background (3), the
$D^0 D^-$ mass spectrum is rescaled according to the ratio of $D^0$
and $D^-$ reconstruction efficiencies and an isospin factor of 1/2.
When this is done, the contribution from background (3) is found to be
negligibly small. Uncertainties in this estimate are included in the
systematic error. Background (4) contributes only to the \ddb\ final
state. It is estimated using a MC simulation of \eeddstng, followed by
\dg.  To reproduce the shape of the \ddst\ mass distribution we use
the $D^{\pm}D{}^{*\mp}$ cross section measured in our previous
study~\cite{belle:dst}.  The contribution from background (4) is found
to be small (shown in Fig.~\ref{fig2}\,(a) as a solid histogram) and
is subtracted from the \ddb\ mass spectrum. Uncertainties in this
estimate are included in the systematic error.  The contribution from
background (5), determined from reconstructed \misid\ events in the
data, is found to be negligibly small and taken into account in the
systematic error.

\begin{figure}[htb]
\hspace*{-0.025\textwidth}
\includegraphics[width=0.5\textwidth]{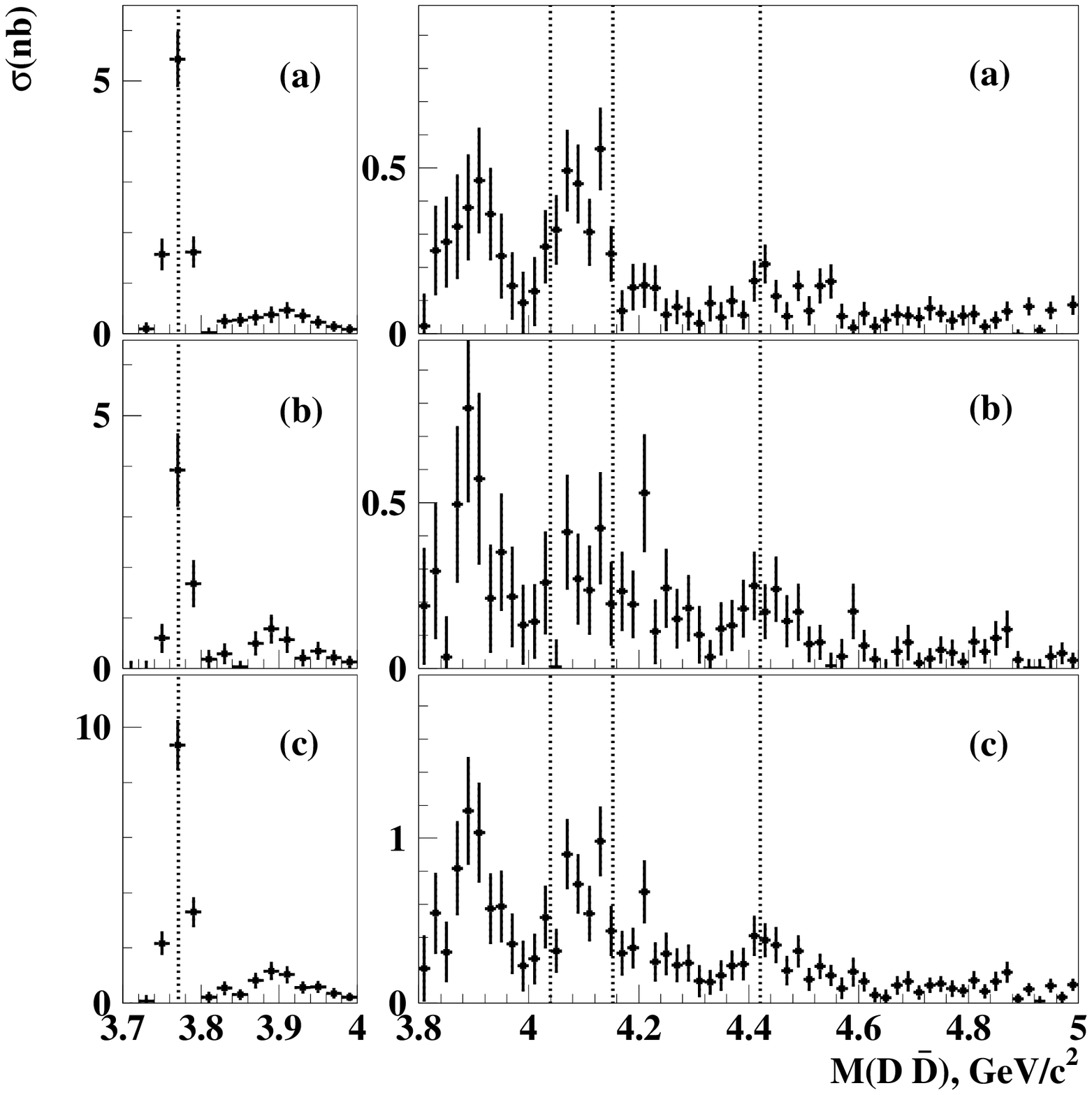}
\caption{The exclusive cross sections for: (a) \eeddb; (b)\eedpdm; (c)
  \eedd.  The dotted lines correspond to the $\psi(3770)$,
  $\psi(4040)$, $\psi(4160)$ and $\psi(4415)$ masses~\cite{pdg}.}
\label{fig3}
\end{figure}

The \eedd\ cross sections are extracted from the \ddb\ and \dpdm\ mass
distributions~\cite{cs}
\begin{eqnarray}
\sigma(\eedd) = \frac{ dN/dm }{ \eta_{\mathrm{tot}} dL/dm} \, ,
\end{eqnarray}
where $m\equiv \mdd$, $dN/dm$ is the obtained mass spectra, while
$\eta_{\mathrm{tot}}$ is the total efficiency. The factor $dL/dm$ is
the differential ISR luminosity
\begin{eqnarray}
dL/dm =\frac{\alpha}{\pi
x}\Bigl((2-2x+x^2)\ln\frac{1+C}{1-C}-x^2C\Bigr) \frac{2m
\mathcal{L}}{E^2_{\mathrm{c.m.}}} \, ,
\end{eqnarray}
where $x = 1 - m^2/E^2_{\mathrm{c.m.}}$, $\mathcal{L}$ is the total
integrated luminosity and $C = \cos\theta_0$, where $\theta_0$ defines
the polar angle range for the \gisr\ in the \ee\ $\mathrm{c.m.}$
frame: $\theta_0<\theta_{\gisr}<180^\circ-\theta_0$.  The total
efficiency determined by MC simulation grows linearly with \mdd\ from
$0.095\%$ near threshold to $0.46\%$ at $5\gevc$ for the \ddb\ and
from $0.038\%$ to $0.17\%$ for the \dpdm\ mode.
The resulting \eeddb, \eedpdm\ and \eedd\ exclusive
cross sections, averaged over the bin width, are shown in
Fig.~\ref{fig3} with statistical uncertainties only.
Since the bin width in the cross section distributions is much larger
than the resolution (which is $\sim 3 \mevc$ at threshold and $\sim 5
\mevc$ at $\mdd \sim 5 \gevc$), no correction for resolution is
applied.

We calculate the cross section ratio ${\seedpdm}/{\seeddb}$ for the
\mdd\ bin $(3.76-3.78)\gevc$ corresponding to $\mdd \approx M_{\ps}$
to be $(0.72\pm 0.16\pm 0.06)$.  This value is in agreement within
errors with CLEO-c~\cite{cleo:dd} and BES~\cite{bes:dd}
measurements. The ratio ${\seedpdm}/{\seeddb}$ integrated over the
\mdd\ range from 3.8 to 5.0$\gevc$ is found to be $ (1.15 \pm 0.13 \pm
0.10)$ and is consistent with unity.

The systematic errors for the \seedd\ measurements are summarized in
Table ~\ref{tab1}.
\begin{table}[htb]
\caption{Contributions to the systematic error in the cross sections,
[$\%$].} \label{tab1}
\begin{center}
\begin{tabular}{@{\hspace{0.3cm}}l@{\hspace{0.3cm}}||@{\hspace{0.3cm}}c@{\hspace{0.3cm}}|@{\hspace{0.4cm}}c@{\hspace{0.3cm}}|@{\hspace{0.4cm}}c@{\hspace{0.3cm}}}
\hline \hline
Source                     & \ddb\    & \dpdm\   & \dd\ \\
\hline
Background subtraction     & $\pm 4$  &  $\pm 3$ &  $\pm 3$   \\
Reconstruction             & $\pm 7$  &  $\pm 6$ &  $\pm 7$   \\
Cross section calculation  & $\pm 5$  &  $\pm 5$ &  $\pm 5$   \\
$\mathcal{B}(D)$           & $\pm 4$  &  $\pm 6$ &  $\pm 5$   \\
Kaon identification        & $\pm 2$  &  $\pm 2$ &  $\pm 2$   \\
\hline
Total                      & $\pm 10$ & $\pm 10$ &  $\pm 10$  \\
\hline \hline
\end{tabular}
\end{center}
\end{table}
The systematic errors associated with the background (1--2)
subtraction are estimated to be 2\% due to the uncertainty in the
scaling factors for the sideband subtractions. This systematic error
is estimated using fits to the \md\ and \mdb\ distributions with
different signal and background parameterizations. Uncertainties in
backgrounds (3--5) are conservatively estimated to be smaller than 2\%
of the signal in the case of \ddb; these two sources are added
linearly to give 4\% in total. In the \dpdm\ case, backgrounds (3--5)
are estimated using the data and only the uncertainty in the scaling
factor for the subtracted distributions is taken into account.  A
second source of systematic error comes from the uncertainties in
track and photon reconstruction efficiencies, which are 1\% per track,
1.5\% per photon and 5\% per $K^0_S$, respectively.  The systematic
error ascribed to the cross section calculation is estimated to be 5\%
and includes the error on the differential ISR luminosity and the
error from the efficiency fit. Other contributions come from the
uncertainty in the identification efficiency and the absolute $D^0$
and $D^+$ branching fractions~\cite{pdg}.  The total systematic
uncertainties are 10\% and comparable to the statistical errors in the
differential cross section around the $\psi(3770)$ peak; for the other
\mdd\ ranges statistical errors dominate.

In summary, we report measurements of \eeddb\ and \eedpdm\ exclusive
cross sections for \sqs\ near the \ddb\ and \dpdm\ thresholds with
initial-state radiation. The observed \eedd\ exclusive cross sections
are consistent with recent BaBar measurements~\cite{babar} and are in
qualitative agreement with the coupled-channel model predictions of
Ref.~\cite{eichten}. This includes the peak at $3.9\gevc$ that is seen
both in Belle and BaBar cross section spectra.

\vspace{0.3cm}

We thank the KEKB group for the excellent operation of the
accelerator, the KEK cryogenics group for the efficient operation of
the solenoid, and the KEK computer group and the National Institute of
Informatics for valuable computing and Super-SINET network support. We
acknowledge support from the Ministry of Education, Culture, Sports,
Science, and Technology of Japan and the Japan Society for the
Promotion of Science; the Australian Research Council and the
Australian Department of Education, Science and Training; the National
Science Foundation of China and the Knowledge Innovation Program of
the Chinese Academy of Sciences under contract No.~10575109 and
IHEP-U-503; the Department of Science and Technology of India; the
BK21 program of the Ministry of Education of Korea, the CHEP SRC
program and Basic Research program (grant No.~R01-2005-000-10089-0) of
the Korea Science and Engineering Foundation, and the Pure Basic
Research Group program of the Korea Research Foundation; the Polish
State Committee for Scientific Research; the Ministry of Education and
Science of the Russian Federation and the Russian Federal Agency for
Atomic Energy; the Slovenian Research Agency; the Swiss National
Science Foundation; the National Science Council and the Ministry of
Education of Taiwan; and the U.S.\ Department of Energy.


\begin{thebibliography} {99}
\bibitem{foot} Charge-conjugate modes are included throughout this paper.
\end{thebibliography}
\end{document}